\documentclass[12pt]{article}
\usepackage{fleqn,epsf,a4wide}

\title{\vspace{-2em} {\hfill\normalsize hep-ph/9907568} \\[-0.4em]
  {\hfill\normalsize DESY 99-093} \\ \vspace{2em}
  Hard exclusive and semi-exclusive \\ meson production\footnote{Talk
    given at the Workshop on the Structure of the Nucleon (NUCLEON
    99), Frascati, Italy, 7--9 June 1999, to appear in the
    proceedings.}}

\author{M. Diehl}

\date{\textit{Deutsches Elektronen-Synchroton DESY, 22603
    Hamburg, Germany}}

\begin{document}

\maketitle

\begin{abstract}
  I review recent theory developments for hard exclusive and
  semi-exclusive production of mesons, emphasising the variety of
  physics issues that can be studied in these processes.
\end{abstract}

\section{EXCLUSIVE MESON PRODUCTION}
\label{sec:excl}

\subsection{Factorisation}
\label{sec:fact}

\begin{figure}
\begin{center}
  \leavevmode
  \epsfxsize 0.97\textwidth
  \epsfbox{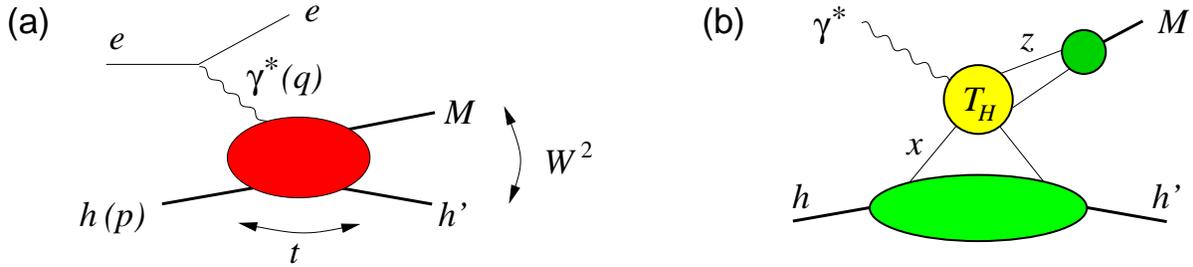}
\end{center}
\caption{{}\label{fig:factorise} (a) Exclusive meson
  electroproduction. (b) Factorisation in the Bjorken limit. $x$ and
  $z$ are parton momentum fractions in $h$ and $M$, respectively.}
\end{figure}

The first part of this review is about exclusive electroproduction of
a meson on a hadron, $e h \to e h' M$, as shown in
Fig.~\ref{fig:factorise}(a). The piece of theory at the origin of the
recent theoretical and experimental interest in this topic is a
factorisation theorem \cite{factoriz-CFS-R}: In the Bjorken limit of
infinitely large photon virtuality $Q^2 = - q^2$ at fixed $x_B = Q^2
/(2 p q)$ and momentum transfer $t$ the amplitude for $\gamma^* h \to
h' M$ factorises into a skewed parton distribution (SPD) in the
target, a hard parton-photon scattering calculable in perturbative
QCD, and the distribution amplitude of the meson, see
Fig.~\ref{fig:factorise}(b). The factorisation property makes this
process a tool to measure leading-twist matrix elements of hadrons, in
particular skewed parton distributions, which have aroused
considerable interest, cf.\ \cite{Rad}.

This factorisation holds only for longitudinal polarisation of the
virtual photon. As $Q^2$ becomes large (and $t$ remains small) the
amplitude from the diagrams in Fig.~\ref{fig:factorise}(b) scales
like $\mathcal{A}(\gamma^*_L) \sim 1/Q$ up to logarithmic corrections.
For transverse photons factorisation cannot be established because the
corresponding diagrams contain dangerous infrared regions, but the
theorem tells us that the amplitude should be suppressed by at least
one power: $\mathcal{A}(\gamma^*_T) \sim 1/Q^2$. Of course, there are
also power corrections to the leading behaviour of
$\mathcal{A}(\gamma^*_L)$.

In the expression of the amplitude one has to perform an integral over
the longitudinal momentum fractions $x$ and $z$ of the partons. An
important point (to which we will come back in
Sect.~\ref{sec:polaris}) is that in order to obtain the leading power
behaviour of the diagrams in Fig.~\ref{fig:factorise}(b) the relative
transverse momenta $k_T$ of the partons within their parent hadrons
are approximated by zero in the hard scattering $T_H$.  Therefore the
SPDs and the meson distribution amplitude are themselves integrated
over $k_T$. Taking into account the effect of finite $k_T$ in $T_H$ is
part of the power corrections to the amplitude.

Important theoretical work has been done in the context of the
high-energy limit, where $W^2$ is much larger than all other
variables, including $Q^2$. Many aspects of it are common with what we
will discuss here, but others are specific to the small-$x_B$ limit
(such as the use of $k_T$-factorisation where the finite parton $k_T$
is not neglected in the hard scattering subprocess). For reasons of
space/time I will not cover this field here.

\subsection{Flavour}

Meson production comes in a variety of different channels, such as
\begin{displaymath}
\renewcommand{\arraystretch}{1.3}
\begin{array}{lcll}
\gamma^* p & \to & 
p\, (\pi^0, \eta, \eta', K^0, \overline{K}^0, \ldots)
& \mbox{(pseudoscalar mesons),} \\
\gamma^* p & \to & p\, (\rho, \rho', \omega, \phi, \ldots)
& \mbox{(vector mesons),} \\
\gamma^* p & \to & n\, (\pi^+, \rho^+, \ldots) , ~
\gamma^* n \:\to\: p\, (\pi^-, \rho^-, \ldots) ~
& \mbox{(charge exchange),} \\
\gamma^* p & \to & (\Lambda, \Sigma^0)\, K^+
& \mbox{(exchange of strangeness).}
\end{array}
\renewcommand{\arraystretch}{1.0}
\end{displaymath}
The multitude offers us a way to disentangle the SPDs for different
parton species and with different spin structure.  Whereas the
distributions $H$ and $E$ (in Ji's notation, corresponding to
Radyushkin's $\mathcal{F}$ and $\mathcal{K}$) occur in vector meson
production their parton helicity dependent counterparts $\tilde{H}$
and $\tilde{E}$ (resp.\ $\mathcal{G}$ and $\mathcal{P}$) contribute to
the production of pseudoscalars. Note that all four distributions
appear in the amplitude for deep virtual Compton scattering, cf.\ 
\cite{Guichon}.

As for the different parton species, one has for instance the
following separation in the vector meson channel \cite{MPW}:
\begin{displaymath}
\renewcommand{\arraystretch}{1.4}
\begin{array}{lcl}
\rho^0 & \sim & {\textstyle \frac{2}{3}} (u + \overline{u}) \,+\,
                {\textstyle \frac{1}{3}} (d + \overline{d}) \,+\,
                \mbox{gluons} , \\
\omega^0 & \sim & {\textstyle \frac{2}{3}} (u + \overline{u}) \,-\,
                  {\textstyle \frac{1}{3}} (d + \overline{d}) \,+\,
                  \mbox{gluons} , \\
\phi & \sim & {\textstyle \frac{1}{3}} (s + \overline{s}) \,+\,
              \mbox{gluons} , \\
\rho^\pm & \sim & (\ldots)\, \left[ (u - \overline{u}) - 
                                    (d - \overline{d}) \right] \,\pm\,
                  (\ldots)\, \left[ (u + \overline{u}) - 
                                    (d + \overline{d}) \right]
                  \hspace{2em} \mbox{(no gluons)} ,
\end{array}
\renewcommand{\arraystretch}{1.0}
\end{displaymath}
where the $(\ldots)$ stand for different $x$-dependent coefficients.
For the charged mesons one has here made use of isospin relations to
relate the SPDs for the transitions $p\to n$ and $n\to p$ to the ones
for $p\to p$ (just as one relates the distribution amplitude of
$\pi^0$, $\pi^-$ and $\pi^+$). A similar separation of flavours can be
achieved in the pseudoscalar channel with pions and kaons \cite{FPPS}.

\begin{figure}
\begin{center}
  \leavevmode
  \epsfxsize 0.97\textwidth
  \epsfbox{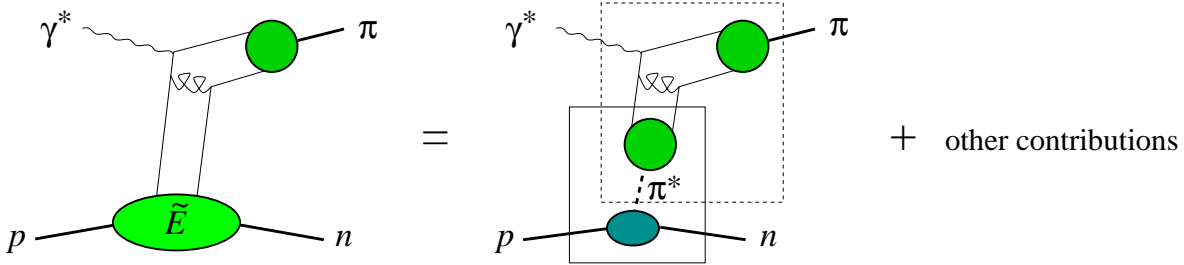}
\end{center}
\caption{{}\label{fig:pion} $\gamma^* p \to n\, \pi^+$: the lower frame
  shows pion exchange as a contribution to the skewed parton
  distribution $\tilde{E}$, the upper frame the (off-shell) pion form
  factor.}
\end{figure}

SPDs contain nonperturbative physics that cannot be accessed in the
usual, diagonal parton distributions. An example is the exchange of
particles in the $t$-channel, which can contribute in the region of
$x$ where both partons come out of the hadron $h$, instead of one
coming out of $h$ and the other going into $h'$ (cf.\ \cite{Rad}).
This is especially important for the distributions $\tilde{E}$, which
have the quantum numbers of pseudoscalar meson exchange: if $t$ is not
too large then an exchanged pion (and to a lesser extent a kaon) are
not very far from their mass shell \cite{FPPS,MPR}. This consideration
has important consequences for the process $\gamma^* p \to n \pi^+$,
which has been used to extract the pion form factor at large spacelike
momentum transfer, cf.\ Fig.~\ref{fig:pion}. Identifying the off-shell
pion in the $t$-channel as a part of the nonperturbative quantity
$\tilde{E}$ it is immediately clear that there are other contributions
to the process, corresponding to contributions in $\tilde{E}$ which
cannot be described by meson exchange. Modelling the SPDs allows one
to estimate the size of such contributions, and to study in which
kinematical region one may hope to extract the pion form factor
\cite{MPR}. This example illustrates how SPDs relate rather different
physics (such as the distribution of quarks in the proton and the
quark-antiquark distribution in in the pion).

A different aspect of pseudoscalar production has been investigated in
\cite{EFS}, where it was proposed to compare the production of
$\pi^0$, $\eta$ and $\eta'$ to study chiral dynamics and the breaking
of flavour $SU(3)$ in the distribution amplitudes of these mesons. The
weighting of $u$-, $d$- and $s$-quarks in electroproduction is
different than from the one in the transitions $\gamma^* \gamma \to
\pi^0, \eta, \eta'$ and thus can provide a handle to separate the
quark flavours.

\subsection{Target Spin}

The cross section for meson production is always quadratic in the
distributions $H$ and $E$ (or $\tilde{H}$ and $\tilde{E}$), even if
the target and recoil hadrons are not polarised. To fully separate the
different distributions one needs to measure the polarisation of at
least one of the hadrons. A kinematic, Rosenbluth-type separation as
can be done for the elastic form factors $F_1$ and $F_2$ is not
possible here, because the functions $H$ and $E$ ($\tilde{H}$ and
$\tilde{E}$) are themselves dependent on the energy of the scattering
process, via the skewedness parameter $\zeta$ (cf.\ \cite{Rad}). As an
example of a polarisation observable the transverse asymmetry of the
target or the recoil hadron has been pointed out in \cite{FPPS}. For
pion production it is proportional to the product $\tilde{E} \cdot
\tilde{H}$ and therefore a good candidate to obtain information on
$\tilde{E}$.

\subsection{From $\rho$ to $\pi\pi$}

\begin{figure}
\begin{center}
  \leavevmode
  \epsfxsize 0.85\textwidth
  \epsfbox{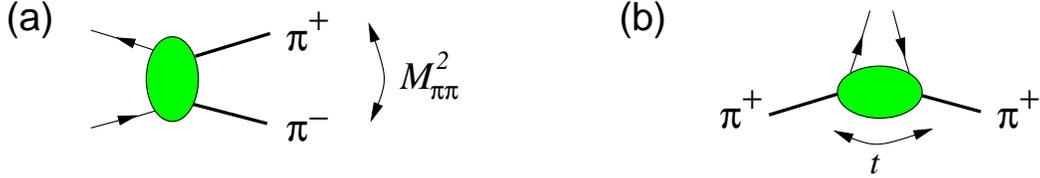}
\end{center}
\caption{{}\label{fig:crossing} (a) Generalised distribution amplitude
  of a pion pair. (b) Parton distribution in a pion, related to the
  generalised distribution amplitude through crossing.}
\end{figure}

So far we have looked at the reaction $\gamma^* p \to p\, \rho^0 \to p
\, (\pi^+\pi^-)$ as the production of a $\rho$, described in the
factorised form of Fig.\ \ref{fig:factorise}(b), followed by the
decay $\rho\to \pi^+\pi^-$. It is however possible to directly
describe the reaction $\gamma^* p \to p \, (\pi^+\pi^-)$ in a
factorised framework, without referring to the formation and decay of
a resonance: the distribution amplitude of the $\rho$ in
Fig.~\ref{fig:factorise}(b) is then replaced with a generalised
distribution amplitude (GDA) \cite{GDA-old,GDA} describing the
transition from the $q \bar{q}$-pair to $\pi^+\pi^-$, cf.\ 
Fig.~\ref{fig:crossing}(a). This was noticed in \cite{PW}, and the
formal extension of the factorisation proof was given in
\cite{factoriz-F}.  It is very natural that this should be possible:
the hadronisation of a $q\bar{q}$-pair into $\pi^+\pi^-$, be it
through resonance formation or not, is all long distance physics and
can be put into one nonperturbative quantity. This tells us that the
study of SPDs in this reaction does not require separation of a
resonance ``signal'' from a continuum ``background'', and is in fact
not restricted to pion pairs on a resonance peak.

A GDA depends on several variables: $(a)$ the momentum fraction $z$ of
the quark with respect to the total momentum, just as for an ordinary
distribution amplitude, $(b)$ the factorisation scale $\mu$---this
dependence is given by the same ERBL evolution equations describing
ordinary distribution amplitudes, $(c)$ the invariant mass
$M_{\pi\pi}$ of the pion pair, and $(d)$ the polar angle $\theta^*$ of
the $\pi^+$ in the $\pi^+\pi^-$ c.m.

The dependence on $\theta^*$ can be described via a decomposition into
partial waves of the $\pi^+\pi^-$ system, and one has two distinct
quantum number combinations: Even partial waves $l = 0, 2, 4, \ldots$
correspond to states with positive charge conjugation parity $C=+1$,
having the quantum numbers of $f$-mesons, and the odd ones, $l = 1, 3,
\ldots$ to $C=-1$, i.e.\ to $\rho$ quantum numbers.  In the
$\rho$-channel the asymptotic form of the GDA, i.e.\ the form one has
at very large factorisation scale $\mu$, has its mass dependence given
by the timelike pion form factor
$F^{\phantom{2}}_{\pi}(M_{\pi\pi}^2)$, which is well measured from
$e^+e^-\to \pi^+\pi^-$ \cite{GDA-P}. This nicely illustrates that,
even in the asymptotic regime, there is more than the $\rho$-resonance
produced---the pion form factor is not described by a Breit-Wigner
form for the $\rho$ alone.

In electroproduction pion pairs in both channels can be produced ,
which implies that there is always some ``contamination'' of
$\rho$-production with the ``wrong'' quantum numbers.  The two
channels go with different flavour combinations of the SPDs:
\begin{displaymath}
\renewcommand{\arraystretch}{1.3}
\begin{array}{lcl}
\rho\mbox{-channel} & \sim &
                      {\textstyle \frac{2}{3}} (u + \overline{u}) \,+\,
                      {\textstyle \frac{1}{3}} (d + \overline{d}) \,+\,
                      \mbox{gluons} , \\
f\mbox{-channel} & \sim &
                      {\textstyle \frac{2}{3}} (u - \overline{u}) \,-\,
                      {\textstyle \frac{1}{3}} (d - \overline{d})
                      \hspace{1.5em} \mbox{(no gluons)} .
\end{array}
\renewcommand{\arraystretch}{1.0}
\end{displaymath}
With a number of simplifying assumptions one can model the GDAs in
both channels, and finds that while on the $\rho$ mass peak the
$\rho$-channel is clearly dominant, the $f$-channel contribution may
be visible if $M_{\pi\pi}$ is a few 100 MeV off peak, provided that
$x_B$ is in the valence region so that gluon exchange does not
completely dominate over quarks \cite{JLab-Procs}. Observable
signatures of the presence of $f$-channel pairs are on one hand a
modified angular distribution of the pion pair compared with a pure
$P$-wave, and in particular the presence of interference terms which
are odd under the exchange of $\pi^+$ and $\pi^-$ momenta. Another
signature of $f$-channel pairs is the production of $\pi^0\pi^0$-pairs
along that of charged pions.

Beyond providing an extended description of meson electroproduction
the GDAs are of interest in themselves, notably because they are
related by crossing symmetry to the (skewed and ordinary) parton
distributions in the pion, cf.\ Fig.~\ref{fig:crossing}. The two types
of nonperturbative quantities can thus be related by analytic
continuation in the invariant mass variable $M_{\pi\pi}$, resp.\ $t$
\cite{GDA-P}. Finally it is clear that the concept of GDAs is not
limited to the $\pi\pi$ system, but can be extended e.g.\ to kaon
pairs, to $p\bar{p}$ or to three pions.

\subsection{Meson Polarisation}
\label{sec:polaris}

As mentioned in sect.~\ref{sec:fact} the factorisation theorem states
that at sufficiently large $Q^2$ longitudinally polarised photons
dominate over transverse ones, i.e.\ it provides us with a helicity
selection rule in the scaling limit. This is important since the
measurement of the azimuthal angle between the electron and hadron
planes in $e h \to e h' M$ contains information about the relative
importance of $\gamma^*_L$ and $\gamma^*_T$ contributions, i.e.\ 
information on how important power corrections are, in other words how
far one is from the asymptotic limit at finite $Q^2$.

It turns out that for vector meson production there is a second
selection rule, stating that to leading power in $1/Q$ only
longitudinally polarised mesons are produced \cite{MPW,DGP}. It easily
generalises to the production of meson pairs: to leading power
accuracy the pair has zero angular momentum along its direction of
flight in the $\gamma^* p$ c.m. Again these selection rules can be
tested by measuring angular distributions.

In the existing data on $\rho$-production, both at fixed-target and at
collider energies, the transition $\gamma^*_T\, p \to p\,
\rho^{\phantom{*}}_T$ is important: even for values of $Q^2$ of 5 to
10 GeV$^2$ the ratio of longitudinal to transverse polarisation is
about 2 to 4 at cross section level \cite{rho-expt}. The H1 and ZEUS
collaborations have also analysed the full angular distributions in
$ep\to ep\,(\pi^+\pi^-)$ and found a hierarchy
$|\mathcal{A}(\gamma^*_L \to \rho^{\phantom{*}}_L)| >$
$|\mathcal{A}(\gamma^*_T \to \rho^{\phantom{*}}_T)| >$
$|\mathcal{A}(\gamma^*_T \to \rho^{\phantom{*}}_L)|$ for those
transition amplitudes that are non-zero within experimental accuracy.

Power corrections are thus seen to be important, certainly in
$\rho$-production. The physics of these corrections contains several
important ingredients. To date most theoretical work has focused on
the effect of finite transverse parton momentum in the hard scattering
subprocess. In the high-energy regime several different estimates of
this give a fair description of the present HERA data on the photon
and $\rho$ polarisation \cite{rho-angles}. Note that meson wave
functions for finite transverse momentum are related by gauge
invariance to wave functions of higher Fock states such as $q\bar{q}
g$ \cite{H-H}. Higher Fock states may thus also play an important
role, but no detailed estimation of this has been performed so far.

In the loop integral of the diagrams in Fig.~\ref{fig:factorise}(b)
there is a region where the momentum fraction $z$ of the quark in the
meson is close to 0 or 1, so that the quark or antiquark becomes slow
and therefore soft. It has long been realised \cite{SJB-et-al} that
for transverse initial photons such infrared sensitive configurations
are not sufficiently suppressed; this is why the simple factorisation
of Fig.~\ref{fig:factorise} cannot be established for a $\gamma^*_T$
\cite{factoriz-CFS-R}. There is an ongoing debate on how serious an
infrared-dependence this introduces in the calculation and to what
extent meson production from a $\gamma^*_T$ can be described in a
perturbative framework, cf.\ for instance \cite{MRT,MP}.

Closely related with this problem is the question how important soft
overlap contributions are relative to the factorising, hard scattering
diagrams. Soft overlap contributions can be obtained from diagrams
such as in Fig.~\ref{fig:pion} by removing the gluon when the quark
line that directly connects the meson and the proton is soft. This is
the same physics that is being intensively discussed for elastic form
factors and other hard exclusive processes \cite{Kroll}.  A first
attempt to estimate the soft overlap contribution to meson
electroproduction has been made in \cite{VGG}.

The theory of power corrections is still far from being complete or
uncontroversial, and it is important to realise that the various
helicity transitions from the photon to the meson (or mesons) provide
a wealth of observables where such effects can be studied without the
need to subtract a leading-twist contribution.

\subsection{Meson production versus Compton scattering}

At this point we have considered several aspects of exclusive meson
production, and it might be useful to confront this reaction with deep
virtual Compton scattering (DVCS) \cite{Guichon}, looking at which
chances and difficulties each of them presents. It comes out that both
processes are very complementary:

\begin{list}{$\bullet$}{\renewcommand{\leftmargin}{1.2em}
    \renewcommand{\itemsep}{0em} \renewcommand{\parsep}{0em}}
\item Unlike Compton scattering we always have two unknown
  non-perturbative quantities in meson production, namely the SPDs in
  the target and the meson distribution amplitude (or GDA).  While the
  latter do not contribute to the $x_B$-dependence of the cross
  section, which is only sensitive to the SPDs, at least the overall
  normalisation of the two quantities cannot easily be disentangled.
  Independent information on meson distribution amplitudes is mainly
  available for the light pseudoscalars, thanks to the measurement of
  the $\gamma$--$\pi$, $\gamma$--$\eta$ and $\gamma$--$\eta'$
  transition form factors.
\item On the other hand, meson production provides a variety of
  channels, which is extremely helpful to disentangle the various
  quantities in their flavour and also in their spin decomposition.
\item Compton scattering offers the unique possibility to measure the
  process at the \emph{amplitude} level, through its interference with
  the Bethe-Heitler process \cite{DGPR}. Extracting this information
  will not be easy, but it may provide the best help one can get on
  the way towards a deconvolution of the $x$-dependence of the SPDs
  from the measured cross sections. Notice in this context the
  different ways in which the SPDs enter in the real and imaginary
  parts of the $\gamma^* p \to \gamma p$ amplitude \cite{Rad,Guichon}.
\item Mechanisms for power corrections are partly different in meson
  production and Compton scattering: in DVCS there is no counterpart
  to corrections due to the $k_T$ in the meson wave function or to
  soft overlap contributions, while in meson production one need not
  worry about the importance of the hadronic, vector-meson type
  component of the produced photon in DVCS. Expectations are that DVCS
  may attain the scaling regime for lower values of $Q^2$ than meson
  production processes. Notice also that DVCS is closely related to
  inclusive deep elastic scattering, represented by the same handbag
  diagrams with the real photon replaced by a virtual one. The
  comparison of the manifestations of higher-twist physics in all
  these reactions may ultimately help us toward understanding of power
  corrections.
\end{list}

\subsection{Proton dissociation}

In addition to quasielastic production $\gamma^* N \to N+M$ one can
consider processes where the nucleon $N$ is excited, e.g.\ into a
nucleon resonance $N^*$ or a $\Delta$, or a $N \pi$ continuum state
with low invariant mass. Being on one hand a background to the
quasielastic process these reactions are interesting in themselves:
factorisation remains valid as long as the meson (or meson pairs) in
the upper part of Fig.~\ref{fig:factorise} and the final state hadrons
from the lower part are well separated in phase space. Then one has
access to SPDs describing transitions such as $N \to \Delta$ or $N \to
N \pi$, on which only little is known so far \cite{FPS}.

The question what happens when one replaces the final state nucleon
with a continuum state whose invariant mass is \emph{not} small leads
us to the second part of this review.

\section{SEMI-EXCLUSIVE MESON PRODUCTION}

\subsection{From exclusive to semi-exclusive}

Let us now look at the process $\gamma^* p \to Y+M$, where $Y$ is a
hadronic system with invariant mass $M_Y$ above the resonance region,
and let us keep the requirement that $Y$ and $M$ be well separated in
phase space. Going through the kinematics of the diagram in
Fig.~\ref{fig:factorise}(b) one finds that for $M_Y \gg m_p$ the
fractional parton momentum $x$, which was previously to be integrated
over all its possible range, becomes ``trapped'' at a particular value
(otherwise one the quarks attached to the soft proton blob in the
diagram must go far off-shell). Moreover, it becomes trapped at a
value for which the photon-parton subdiagrams $T_H$ have internal
lines close to their mass-shell. This means that the ``hard
scattering'' is no longer hard, and that to factorise diagrams as it
is done in Fig.~\ref{fig:factorise}(b) is no longer the correct way to
describe the process.

A way out of this situation is to require a large invariant momentum
transfer $t$ between the proton and $Y$, i.e.\ between the photon and
$M$. Then the value of $x$ where the integration is ``trapped'' no
longer corresponds to a singularity of the hard scattering process,
which remains under perturbative control. Once we have large $t$ we
can allow $Q^2$ to be small or zero (and even replace the photon
projectile with a hadron, say a pion). This defines then the
``semi-exclusive'' kinematics introduced and studied in \cite{BDHP}:
\begin{list}{$\bullet$}{\renewcommand{\leftmargin}{1.2em}
    \renewcommand{\itemsep}{0em}}
\item $-t, M_Y^2 \gg 1~\mbox{GeV}^2$,
\item $-t, M_Y^2 \ll W^2$ so that $Y$ and $M$ are well separated,
\item $Q^2$ zero, small or large.
\end{list}

\begin{figure}[b]
\begin{center}
  \leavevmode
  \epsfxsize 0.97\textwidth
  \epsfbox{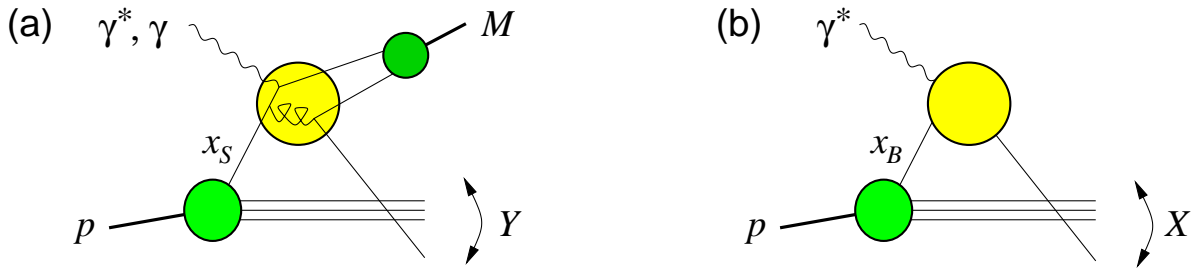}
\end{center}
\caption{{}\label{fig:semi-excl} (a) Factorisation in semi-exclusive
  meson production, $\gamma^{(*)} p \to Y+M$, with an example for the
  hard-scattering subdiagrams. (b)~Inclusive DIS, $\gamma^* p \to X$.}
\end{figure}

The process can then be calculated in a factorised way, shown in
Fig.~\ref{fig:semi-excl}(a). The upper part of the diagram is the same
as for exclusive meson production in the Bjorken region, composed of a
hard parton-photon scattering and a meson distribution amplitude (or
GDA), while the lower part has changed. The reason is that the parton
``coming back'' from the hard scattering has a large transverse
momentum compared to the spectator partons in the proton, and
therefore hadronises independently from the spectators. This is
precisely the same situation as in inclusive deep inelastic scattering
(DIS), cf.\ Fig.~\ref{fig:semi-excl}(b), so that in the calculation we
can treat the scattered parton as a free final-state particle. In the
squared transition amplitude the parton emission from the target is
just described by an ordinary (non-skewed) parton distribution. If the
target is polarised then one probes the usual helicity-dependent
parton distributions.

Reminding ourselves that in DIS we have a hard scattering on a parton
with momentum fraction $x_B = Q^2 /(W^2 + Q^2)$ and making the
necessary translation of variables we find that in our semi-exclusive
reaction the extracted parton has momentum fraction $x_S = (-t)
/(M_Y^2 - t)$. The analogy with DIS also indicates that one may invoke
parton-hadron duality to use the description of
Fig.~\ref{fig:semi-excl}(a) even when the mass $M_Y$ is in the
resonance region, on the condition of integrating over a sufficiently
large $M_Y$-range so that individual resonances average out.

At first sight it may appear that our process is just a special case
of a semi-\emph{inclusive} reaction, the formation of the meson $M$
being described by an ordinary fragmentation function with momentum
fraction $z$ close to 1. The example diagram in
Fig.~\ref{fig:semi-excl}(a) shows that this is not the case: we do
not have there a two-step process where a quark from the proton is
first scattered off the photon, and then fragments into the meson $M$
and some other partons. Taking suitable kinematical limits one can
however find situations where the close connection between the
semi-exclusive and semi-inclusive mechanisms becomes apparent; such
cases have in fact been considered in the literature \cite{BKM,CW}.

\subsection{What one can learn in semi-exclusive processes}

Semi-exclusive processes have several features that make them
interesting. Let me just mention, but not elaborate on the possibility
of investigating large-$t$ Regge exchange in the context of
perturbative QCD. Vector meson production is already being extensively
studied under this aspect, both theoretically and experimentally (cf.\
\cite{BDHP} for references).

Semi-exclusive reactions have a hybrid nature: the upper part of the
diagram in Fig.~\ref{fig:semi-excl}(a) is typical of exclusive
reactions, and offers a way to study meson distribution amplitudes. It
may be particularly attractive to consider ratios of cross sections,
where many details of (and corrections to) the calculation cancel out.
The cross section ratio $d \sigma(\rho^+) /d \sigma(\pi^+)$ for
instance only depends on the distribution amplitudes $\phi_\rho(z)$
and $\phi_\pi(z)$.

The lower part of the same diagram is reminiscent of usual inclusive
or semi-inclusive reactions; the cross section of our process is
linear in the ordinary parton distributions, not quadratic in skewed
distributions as the exclusive processes we have discussed in
Sect.~\ref{sec:excl}. Producing different mesons one can select
particular flavour combinations of parton distributions (in a similar
way as in semi-inclusive processes, but without the uncertainties due
to ``unfavoured'' fragmentation functions). Examples are the ratios
\begin{displaymath}
  \frac{d \sigma(\gamma p \to Y \pi^-)}{
        d \sigma(\gamma p \to Y \pi^+)} = 
     \frac{d(x_S) + \overline{u}(x_S)}{u(x_S) + \overline{d}(x_S)} ,
\hspace{3em}
  \frac{d \sigma(\gamma p \to Y K^-)}{
        d \sigma(\gamma p \to Y K^+)} =
     \frac{s(x_S) + \overline{u}(x_S)}{u(x_S) + \overline{s}(x_S)} ,
\end{displaymath}
and those with $\Delta d(x_S), \Delta \overline{u}(x_S)$, \ldots\ one
can measure with longitudinal target polarisation. Note that the
comparison of $d(x)$ with $u(x)$, or $\overline{s}(x)$ with $s(x)$
provides valuable information on non-perturbative dynamics in the
proton.

To obtain a feeling for accessible values of $x_S$ let us consider
some example kinematics for $ep$ c.m.\ energies $\sqrt{s}$ attainable
at ELFE and at COMPASS:
\begin{center}
$
\renewcommand{\arraystretch}{1.2}
\begin{array}{ccccc} \hline
\sqrt{s}\; [\mbox{GeV}] & W\; [\mbox{GeV}] &
M_Y^2\; [\mbox{GeV}^2] & -t\; [\mbox{GeV}^2] &
x_S = (-t) /(M_Y^2 - t) \\ \hline
 8 &  6 &  2 & 6 & 0.75 \\
   &    &  6 & 2 & 0.25 \\
20 & 15 & 20 & 3 & 0.13 \\
   &    & 25 & 2 & 0.07 \\ \hline
\end{array}
\renewcommand{\arraystretch}{1.0}
$
\end{center}
At ELFE with $\sqrt{s} \approx 8$ GeV the range of masses $M_Y$ for
which $Y$ and the meson are well separated is rather limited, and one
always remains in the valence region of the parton distributions. On
the other hand, high luminosity may allow one to go to rather high $t$
and thus to access parton distributions at large $x$. For COMPASS
energies there is a comfortable range of $W$ and hence of $M_Y$, and
one would be able to go down to the $x$-region where sea quarks and
gluons are important.

\section{SUMMARY}

Exclusive production of mesons in the Bjorken region provides an
opportunity to study meson distribution amplitudes and skewed parton
distributions, and is in several ways complementary to deep virtual
Compton scattering. It offers a large variety of channels and thus the
possibility to disentangle the many flavour degrees of freedom. Some
channels have (unexpected) connections to other processes, for
instance $\pi^+$ production with its relation to the elastic pion form
factor.

The spin degrees of freedom of SPDs can partly be disentangled by
comparing different meson channels and DVCS, but full information can
only be obtained if the polarisation of the target and/or recoil
hadron is measured. The polarisations of the virtual photon and of the
produced meson play a rather different role: they are indicators of
how important power corrections are. This should help us to judge when
a leading-twist interpretation of the data is adequate, and beyond
this to learn more about non-leading twist physics.

Generalised distribution amplitudes open further channels to a
factorised description within QCD, with low-mass meson pairs instead
of single mesons. They provide a new way to think about and analyse
the production of unstable resonances such as the $\rho$. By crossing
they are connected to parton distributions in mesons and thus allow
one to relate different pieces of information on hadron structure.

Semi-exclusive production combines typical features of exclusive and
inclusive physics. Next to the possibility to study Regge physics in
the perturbative region it offers ways to compare different meson
distribution amplitudes, and to measure ``exotic'' flavour
combinations of unpolarised and polarised parton distributions.

Clearly there remains much theoretical work to be done on several of
these issues, and much experimental work, too. The physics potential I
have tried to outline will hopefully make these efforts worthwhile.


\begin{thebibliography}{99}
  
\bibitem{factoriz-CFS-R} J.C. Collins, L. Frankfurt and M. Strikman,
  Phys.\ Rev.\ D56, 2982 (1997), hep-ph/9611433;\\
  A.V. Radyushkin, Phys.\ Rev.\ D56, 5524 (1997), hep-ph/9704207.

\bibitem{Rad} A.V. Radyushkin, these proceedings.
  
\bibitem{Guichon} P.A.M. Guichon, these proceedings.

\bibitem{MPW} L. Mankiewicz, G. Piller and T. Weigl, Eur.\ Phys.\ J.
  C5, 119 (1998), hep-ph/9711227
  and Phys.\ Rev.\ D59, 017501 (1999), hep-ph/9712508.

\bibitem{FPPS} L.L. Frankfurt, P.V. Pobylitsa, M.V. Polyakov and M.
  Strikman, Phys.\ Rev.\ D60, 014010 (1999), hep-ph/9901429.

\bibitem{MPR} L. Mankiewicz, G. Piller and A. Radyushkin,
  hep-ph/9812467.

\bibitem{EFS} M.I. Eides, L.L. Frankfurt and M.I. Strikman, Phys.\ Rev.\
  D59, 114025 (1999), hep-ph/9809277.

\bibitem{GDA-old} D. M\"uller, D. Robaschik, B. Geyer, F.M. Dittes and J.
  Ho\v{r}ej\v{s}i, Fortsch.\ Phys.\ 42, 101 (1994), hep-ph/9812448.
  
\bibitem{GDA} M. Diehl, T. Gousset, B. Pire and O. Teryaev, Phys.\ 
  Rev.\ Lett.\ 81, 1782 (1998), hep-ph/9805380.

\bibitem{PW} M.V. Polyakov and C. Weiss, Phys.\ Rev.\ D59, 091502
  (1999), hep-ph/9806390.

\bibitem{factoriz-F} A. Freund, hep-ph/9903489.
  
\bibitem{GDA-P} M.V. Polyakov, hep-ph/9809483;\\
  M.V. Polyakov and C. Weiss, hep-ph/9902451.
 
\bibitem{JLab-Procs} M. Diehl, T. Gousset and B. Pire, to appear in
  the Procs.\ of the Workshop on Exclusive and Semiexclusive Processes
  at High Momentum Transfer, Jefferson Lab, USA, 20--22 May~1999.

\bibitem{DGP} M. Diehl, T. Gousset and B. Pire, 
Phys.\ Rev.\ D59, 034023 (1999), hep-ph/9808479.

\bibitem{rho-expt} ZEUS Collaboration, Eur.\ Phys.\ J. C6, 603 (1999),
  hep-ex/9808020;\\
  B. Clerbaux (H1 Collaboration), hep-ph/9905507;\\
  A. Borissov (HERMES Collaboration), to appear in the Procs.\ of the
  7th International Workshop on Deep Inelastic Scattering and QCD (DIS
  99), Zeuthen, Germany, 19--23 Apr.~1999.
  
\bibitem{rho-angles} D.Yu.\ Ivanov and R. Kirschner, Phys.\ Rev.\ D58,
  114026 (1998), hep-ph/9807324;\\
  E.V. Kuraev, N.N. Nikolaev and B.G. Zakharov, JETP Lett.\ 68, 696
  (1998), hep-ph/9809539;\\
  I. Royen, to appear in the Procs.\ or the 7th International Workshop
  on Deep Inelastic Scattering and QCD (DIS 99), Zeuthen, Germany,
  19--23 Apr.~1999.
  
\bibitem{H-H} I. Halperin, Phys.\ Rev.\ D57, 1680 (1998),
  hep-ph/9704265;\\
  P. Hoodbhoy, Phys.\ Rev.\ D56, 388 (1997), hep-ph/9611207.

\bibitem{SJB-et-al} S.J. Brodsky et al., Phys.\ Rev.\ D50, 3134
  (1994), hep-ph/9402283.
  
\bibitem{MRT} A.D. Martin, M.G. Ryskin and T. Teubner, Phys.\ Rev.\ 
  D55, 4329 (1997), hep-ph/9609448.

\bibitem{MP} L. Mankiewicz and G. Piller, hep-ph/9905287.

\bibitem{Kroll} P. Kroll, these proceedings.

\bibitem{VGG} M. Vanderhaeghen, P.A.M. Guichon and M. Guidal,
  hep-ph/9905372.
  
\bibitem{DGPR} M. Diehl, T. Gousset, B. Pire and J.P. Ralston, Phys.\ 
  Lett.\ B411, 193 (1997), hep-ph/9706344.

\bibitem{FPS} L.L. Frankfurt, M.V. Polyakov and M. Strikman,
  hep-ph/9808449.
  
\bibitem{BDHP} S.J. Brodsky, M. Diehl, P. Hoyer, S. Peign\'{e}, Phys.\
  Lett.\ B449, 306 (1999), hep-ph/9812277.
  
\bibitem{BKM} A. Brandenburg, V.V. Khoze and D. M\"uller,
  Phys.\ Lett.\ B347, 413 (1995), hep-ph/9410327.
  
\bibitem{CW}  C.E. Carlson and A.B. Wakely, Phys. Rev.\ D48, 2000
  (1993);\\
  A. Afanasev, C.E. Carlson and C. Wahlquist, Phys.\ Lett.\ B398, 393
  (1997), hep-ph/9701215
  and Phys.\ Rev.\ D58, 054007 (1998), hep-ph/9706522.


\end{thebibliography}
\end{document}